\documentclass[12pt,showpacs,preprintnumbers,amsmath,amssymb]{revtex4}

\usepackage{epsf}
\usepackage{graphicx}  % Include figure files
\usepackage{dcolumn}   % Align table columns on decimal point
\usepackage{bm}        % bold math

\newcommand{\be}{\begin{equation}}
\newcommand{\en}{\end{equation}}

\begin{document}

\title{On asymmetric brane creation}
\author{Hongsheng Zhang  \footnote{Electronic address: hongsheng@shnu.edu.cn} }
 \affiliation{ Shanghai United Center for
Astrophysics (SUCA), Shanghai Normal University, 100 Guilin Road,
Shanghai 200234,  China \\
Kavli Institute for Theoretical Physics China (KITPC), Chinese Academy of Science, Beijing,100190, China}
 \author{Zongkuan Guo\footnote{Electronic address:
guozk@itp.ac.cn}}
 \affiliation{State Key Laboratory of Theoretical Physics(SKLTP)
Institute of Theoretical Physics, Chinese Academy of Science, Beijing,100190, China}
\author{Chi-Yi Chen${}^{a}$}
\email{chenchiyi@hznu.edu.cn}
\affiliation{${}^a$Hangzhou  Normal University, Hangzhou 310036, 
China}
\author{Xin-Zhou Li \footnote{Electronic address: kychz@shnu.edu.cn} }
 \affiliation{Shanghai United Center for Astrophysics (SUCA), Shanghai Normal
University, 100 Guilin Road, Shanghai 200234, China}

\begin{abstract}
We exhaust the brane instanton solutions---an Einstein brane
inhabiting at different positions in a 5-dimensional Einstein bulk with negative
curvature. We construct a brane instanton model
consisting of a brane with asymmetric bulk along two sides of the
brane. And the junction condition of the resulting space-time is
analyzed in the frame of induced gravity (DGP model). In spirits of quantum
gravity of path integral formulism we calculate the Euclidean
actions on three canonical paths and then compare the Euclidean
actions of different instantons per unit 4-volume. We also compare
the Euclidean actions per unit 4-volume of instantons, which consist of
a brane gluing to a fixed  half, with other Euclidean actions of
 halves possessing different cosmological constants.

\end{abstract}


\maketitle

\section{Introduction}

 Quantum cosmology deals with the creation of our universe where and when the classical
 gravity theory may fail. In the formula
 \be
 \Psi[\Sigma,h] =\sum_M\int D[g_E]{\rm exp}(-S_E[M,g_E]),
 \label{basic}
 \en
 $\Psi$ is the propagator from a 3-manifold (usually {\it nothing} taken) to
 ($\Sigma,h)$, and $S_E$ is the Euclidean action of the ``path" $M$.
   The standard model of the quantum origin of the universe
begins with a consisting of a compact, path connected, oriented,
Riemannian manifold $M_R$ adhered to a Lorentzian manifold $M_L$
by a totally geodesic spacelike hypersurface $\Sigma$, which serves
as an initial Cauchy surface for the Lorentzian evolution on
manifold $M_L$. Given this setup we pass to the double $2M_R=M_R^z
\cup M_R^y$ by joining two copies of $M_R$ across $\Sigma$. This
is a closed, path connected, oriented, Riemannian 4-manifold $M =
2M_R$ with a mirror isometry that fixes the totally geodesic
submanifold $\Sigma$. Here $M$ is a very generic topological
manifold. The smaller set is often considered : the gravitational
instantons, that is, $M$ is an Einstein manifold \cite{qc}. Here
we generalize the conception of instanton---we permit non mirror
symmetry between $M_R^z$ and $M_R^y$.

  On the other hand the concept brane emerging recent
  years is very important in high energy physics and cosmology.
  In the brane world
scenario, the standard model particles are confined to the
3-brane, while gravity can propagate in the whole space
\cite{braneworld}.  An inflating brane world with positive curvature
and mirror symmetry created from ``nothing'' together with
its Anti de Sitter (AdS) bulk has been considered in
\cite{garriga}.  Then
the creation of the inflationary brane universe in 5d bulk
Einstein and Einstein-Gauss-Bonnet gravity has been analyzed in
\cite{akm}. Brane instantons in F-theory have been explored in \cite{fbrane}.
The relation between brane instanton and Taub-NUT in M-theory is studied in \cite{witten}. Brane instanton intersecting at sine angle is presented in \cite{angle}. Many different types of brane world creation models
have been investigated widely \cite{brcrea}.  The brane world
models without mirror symmetry have been investigated to some
extent: different black hole masses on the two sides on the brane
\cite{non1}, zero black hole mass and different cosmological
constants on the two sides \cite{non2} and allowing for both types
of generalizations \cite{nonboth}. The quantum creation of closed
branes by totally antisymmetric tensor and gravity was treated as
an interesting solution to cosmological problem in \cite{referee}.
In this paper, we shall investigate the creation of a brane world model with non
mirror symmetric bulk.

 We study all the three types of
solutions---negative curvature, positive curvature and Ricci flat
Einstein branes inhabiting at different positions in a 5d negative
curvature Einstein manifold and investigate instantons including
brane without mirror symmetric in bulk in section II. Negative
curvature Einstein manifold means $R=kg$, where $R$ is Ricci
tensor, $g$ denotes metric and $k$ is a negative real number;
while positive curvature is characterized by a positive $k$. In
section III we analyze the junction condition of the brane
instanton solutions in the induced gravity frame. The main purpose
of section IV is to obtain the actions of the different three
types of branes, with or without induced gravity term so as to
finding some clues of comparing their creation probabilities.
Different from the symmetric case \cite{garriga} we find the Gibbons-Hawking
boundary term in context of asymmetric instanton must be
considered. Finally we present our conclusions and discussions in
section V.

\section{instanton solutions}

We consider a 4d brane embedded in a 5d Einstein bulk. In Gaussian
normal coordinates the metric ansatz of a 5d Euclidean-Einstein
space is written as
\begin{equation}
{}^{(5)}g_E= g_{ab}dx^a dx^b= dr^2+b^2(r)ds_4^2.
 \label{mcbulk}
\end{equation}
Here Latin indices run from 0 to 4, $b(r)$ has the dimension of
length. The induced 4-metric on the brane is $g_E=b^2(r)ds_4^2$~.
Then we introduce $h_E=ds_4^2$ for convenience.

 There is an interesting relation between ${}^{(5)}g_E$ and $g_E$, as shown in the following lemma : \\
{\slshape The bulk with metric ansatz (\ref{mcbulk}) is a negative
 curvature Einstein manifold only if it has an Einstein
submanifold.}

 Pf:  One selects an orthonormal base $ e^r=dr, e^{\mu}=b
 \tilde{e^{\mu}}$, where $\tilde{e^{\mu}}$ is an orthonormal base of the
 $h_E$, Greek index labels 0--3. Under this base one has

 \be
 {}^{(5)}R_{rr}=-\frac{4b''}{b},
 \label{riccir}
 \en
 where prime stands for derivative with respect to $r$. If ${}^{(5)}g_E$ is an Einstein
 manifold, one has ${}^{(5)}R_{rr}=C g_{rr}$~, where $C$ is a constant. Set
 $C=-\frac{4}{l^2}$ then one further has

 \be
 l^2b''=b.
 \en
 The general solution of the above equation is
 \be
 b=c_1e^{\frac{r}{l}}+c_2e^{-\frac{r}{l}},
 \label{bfunction}
 \en
where $c_1,~c_2$ are two constants and $l$ has the dimension of
length. Along the directions of the submanifold we find

 \be
 {}^{(5)}R_{\alpha\mu}=\frac{1}{b^2}[\tilde{R}_{\alpha\mu}
 -3b'^2 \delta_{\alpha\mu}]-\frac{1}{l^2}\delta_{\alpha\mu}.
 \label{ricci5}
 \en
  If ${}^{(5)}g_E$ is an Einstein manifold, i.e., ${}^{(5)}R_{\alpha\mu}=-\frac{4}{l^2}
 \delta_{\alpha\mu}$, substituting (\ref{bfunction}) into (\ref{ricci5}),
 one has
 \be
 \frac{1}{(c_1e^{r/l}+c_2e^{-r/l})^2} \left[
 \tilde{R}_{\alpha\mu}- \frac{3}{l^2}\left(
 c_1e^{r/l}-c_2e^{-r/l}\right)^2 \delta_{\alpha\mu}\right]
 -\frac{1}{l^2}\delta_{\alpha\mu}=-\frac{4}{l^2}\delta_{\alpha\mu}~,
 \en
 therefore one obtains
 \be
 \tilde{R}_{\alpha\mu}=-\frac{3}{l^2}\delta_{\alpha\mu}(4c_1c_2)~.~~~\Box
 \label{rtilde}
 \en

 We see this Ricci tensor $\tilde{R}_{\alpha\mu}$ can be treated as a solution of 4d vacuum Einstein
 equation with a cosmological constant $-{4c_1c_2}/{l^2}$. For giving prominence to
 the key point of this paper, we set
 \be
  -{4c_1c_2}/{l^2}=\epsilon,
  \label{normalc}
  \en
  where $\epsilon=-1,~0,~1$ denote
 negative curvature, Ricci flat or positive curvature canonical
 submanifold respectively. In
 this setup we have put the dimension of $g_E$ into $b^2$, so both the components
 of $h_E$ and the coordinates $x^\mu$ are dimensionless. Hence
 $\tilde{R}_{\alpha\mu}$  in (\ref{rtilde}) is dimensionless
 however ${R}_{\alpha\mu}(g_E)$ has dimension of [length]$^{-2}$.
 Simple calculation gives
 \be
 {R}_{\alpha\mu}(g_E)=\epsilon b^{-2}\delta_{\alpha\mu}.
 \en
  $c_1$ and $c_2$ have dimension of [length].

 Here we point out the 3 types of hypersurfaces, i. e., the positive,
  negative and Ricci flat hypersurfaces correspond to different slicings of 5d
  Einstein manifold. To illuminate this point first we
 suppress 3d of the original 5d manifold, taking a cross section of the 5d
 manifold. Then we take a Wick rotation to return back to Lorentz manifold and then choose
 $c_1=-c_2=l/2,~c_1=c_2=l/2$ and $~c_1=l,~c_2=0$ to represent
 canonical positive curvature, negative curvature and Ricci flat submanifold
 respectively. If the space time
 we considered is not maximally  symmetric it can not be imbedded in a 6d flat manifold.
 But we always can imbed it into a Ricci flat 6d manifold \cite{imbed}.
 Either embedded into a 6d flat or 6d Ricci flat spacetime the cross section
 keeps the same.  It is obvious the left 2d manifold
 is maximally symmetric manifold. We imbed this maximally symmetric
 manifold in the following 3d pseudo-Euclidean manifold

 \be
 ds^2=-dT^2-dW^2+dZ^2~,
 \en
 where
 \be
 -T^2-W^2+Z^2=-l^2
 \en
 In the following parametrization
 \be
 T=l\cosh(r_1/l)\sin(t_1)~,
 \nonumber
 \en
 \be
 W=l\cosh(r_1/l)\cos(t_1)~,
 \nonumber
 \en
 \be
 Z=l\sinh(r_1/l),
 \label{1stchar}
 \en
  we have
  \be
  g=dr_1^2-l^2\cosh^2(r_1/l)dt_1^2~.
  \label{neg}
  \en
 This chart covers the whole manifold except some singularities
 which form a zero measurement set. In this chart negative
 hypersurfaces stand at $r_1=$ constant.

 In the parametrization of
 \be
 T=le^{r_2/l}t_2~,
 \nonumber
 \en
 \be
 W=l\cosh(r_2/l)-\frac{1}{2}le^{r_2/l}t_2^2~,
 \nonumber
 \en
 \be
 Z=l\sinh(r_2/l)+\frac{1}{2}le^{r_2/l}t_2^2~,
 \label{2ndchar}
 \en
 the metric becomes
 \be
 g=dr_2^2-l^2e^{2r_2/l}dt_2^2~.
 \label{fla}
 \en
 This chart covers half of the manifold, namely, the region $Z+W>0$.
 In this chart Ricci flat
 hypersurfaces stand at $r_2=$ constant.
 The third parametrization is
 \be
 T=l\sinh(r_3/l)\sinh(t_3)~,
 \nonumber
 \en
 \be
 W=l\cosh(r_3/l)~,
 \nonumber
 \en
 \be
 Z=l\sinh(r_3/l)\cosh(t_3),
 \label{3rdchar}
 \en
 by which we derive the metric
 \be
 g=dr_3^2-l^2\sinh^2(r_3/l)dt_3^2~.
 \label{pos}
 \en
 This chart covers half of the manifold, that is, the region $W>0$.
 In this chart positive curvature
 hypersurfaces stand at $r_3=$ constant.
 Now we present a simple
 conclusion which is not priorly obvious.
 We know there are 3 classes of parameterizations depending on the sign of  $c_1c_2$.
 All the metrics of hypersurfaces in
the same class can be written in standard form, given by
(\ref{neg}), (\ref{fla}) or (\ref{pos}), simply by a coordinate
transformation. Here we prove this conclusion in the positive
curvature class. For any $c_1c_2<0~,$
 define new coordinates $t_5,r_5$ by
 \be
 T=\frac{l}{\sqrt{-4c_1c_2}}(c_1e^{r_5/l}+c_2e^{-r_5/l})\sinh t_5,
 \nonumber
 \en
 \be
 W=\frac{l}{\sqrt{-4c_1c_2}}(c_1e^{r_5/l}-c_2e^{-r_5/l}),
 \nonumber
 \en
 \be
 Z=\frac{l}{\sqrt{-4c_1c_2}}(c_1e^{r_5/l}+c_2e^{-r_5/l})\cosh t_5.
  \en
We find
 \be
 g=dr_5^2-b^2dt_5^2~,
 \en
 where
 \be
 b=\frac{l}{\sqrt{-4c_1c_2}}(c_1e^{r_5/l}+c_2e^{-r_5/l}).
 \label{gb}
 \en
 Considering the normalization condition (\ref{normalc}) it is
 just (\ref{bfunction}).
 It is easy to find  transformation between $(r_5,~t_5)$ and
 $(r_3,~t_3)$,
 \be
 t_3=t_5
 \nonumber
 \en
 \be
 \sinh(r_3/l)=\frac{1}{\sqrt{-4c_1c_2}}(c_1e^{r_5/l}+c_2e^{-r_5/l}).
 \label{tc1c2}
 \en
 Evidently for any $c_1,~c_2$ in the same family, such as positive slicing, with a general
 conformal factor (\ref{gb}) the only work to get the standard
 chart is to rescale the coordinate $r_5$. Similar conclusions
 hold for the cases of negative and Ricci flat slicings. Therefore we always
 suppose that the metric of Einstein bulk together with its Einstein brane
 has been rescaled to the standard form (\ref{neg}), (\ref{fla}) or (\ref{pos}).
 This assumpotion carries many conveniences for the discussions
 of asymmetric instanton solutions.

  If one only considers mirror symmetric instanton one can
  construct it by excising the spacetime region at $r> r_0$ left
$M_R$ and gluing two copies of the remaining spacetime along the
4-hypersurface at $r=r_0$ , $M=M_R^z\bigcup M_R^y$. We consider
the case where the bulk on two sides of the brane have no mirror
symmetry, that is, the cosmological constants are different on the
two sides of the brane. Because the 4-metric induced by the
5-metric of left bulk must be identical to the metric induced by
right bulk, i.e., the 4-metric on the brane is unique, we have
 \be
 g_E^{left}=g_E^{right}.
 \label{jun}
 \en

\begin{figure}
 \centering
 \includegraphics[totalheight=4in]{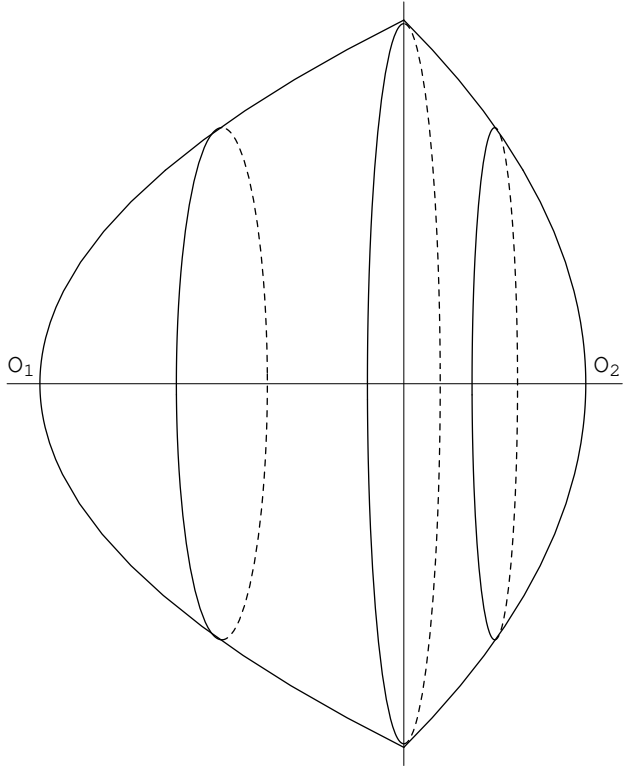}
 \caption{ Outline of an asymmetric instanton. }
 \label{non}
 \end{figure}

   Obviously if $g_E^{left}$ belongs to the positive curvature class
   but $g_E^{right}$ belongs to the negative curvature class the bulk on
   two side cannot be glued together. So we consider to glue two halves
   of the bulk with sectional
 hypersurface in the same canonical class but different cosmological constants.
 For example to the brane in the positive curvature class the junction
  condition (\ref{jun}) becomes
 \be
 l_1 \sinh(r_1/l_1)=l_2 \sinh(r_2/l_2),
 \label{warpj}
 \en
 where $l_1^2=-{6}/{{}^{(5)}\Lambda_L}$ ($l_2^2=-{6}/{{}^{(5)}\Lambda_R}$) is the characteristic
 length of the left (right) bulk and $r_1$ ($r_2$) the position of the brane
 in the left (right) bulk. The
 outline of an asymmetric instanton is shown in fig. \ref{non}.

  Note that the maximally symmetric manifold---Minkowski, de Sitter or
  anti-de Sitter, is certainly Einstein manifold but
  we have many other choices, Schwarzschild (Schwarzschild-(Anti) de Sitter) solution,
   Kerr (Kerr-(Anti) de Sitter) solution etc. Here we note
   Randall-sundrum Minkowski brane is the flat solution included in Ricci flat class with
   $c_1c_2=0,~ c_1^2+c_2^2\neq 0$. Furthermore for convenience of later development
   we give (Anti) de Sitter manifold and Schwarzschild-((Anti) de
   Sitter)
   manifold as $ds_4^2$ clearly in Euclidean form.
   For (Anti) de Sitter
   \be
   {\rm de~Sitter}~~~~~~ds^2_4=d\chi^2+\sin^2\chi d\Omega_{(3)}^2,
   \label{dsbrane}
   \en
   \be
   {\rm anti~ de~Sitter}~~~~~~ds^2_4=d\chi^2+\sinh^2\chi d\Omega_{(3)}^2,
   \label{adsbrane}
   \en
 where $d\Omega_{(3)}^2$ is 3-sphere. For Schwarzschild-((Anti-) de
 Sitter)
 \be
 ds_4^2=(1-\frac{2m}{r'}+\epsilon r'^2)d\chi^2+ (1-\frac{2m}{r'}+
 \epsilon r'^2)^{-1}dr'^2+r'^2d\Omega_{(2)}^2,
 \label{sch(a)ds}
 \en
 where $r'$ is the radial coordinate on the brane and
 $d\Omega_{(2)}^2$ is 2-sphere. Kerr-(Kerr-(Anti) de Sitter) solution and any other
 Einstein manifold can be written similarly.
  \section{junction condition}

 In this section we shall investigate the
 general junction in frame of brane--induced--gravity. The induced gravity brane model
 was proposed by Dvali et al. \cite{Dvali}, and we work in the
 generalized Dvali-Gabadaze-Porrati model presented in \cite{eff}.

 We consider an asymmetric 5d gravitational instanton with a 4d brane on which
  an induced Ricci scalar term is confined,
 \be S=S_{L \rm bulk}+S_{\rm brane}+S_{R \rm bulk},
 \en
where
 \be S_{L \rm bulk} =\int d^5X \sqrt{-{\rm det}({}^{(5)}g_L)}
 \left[ {1 \over {16\pi G_5} }( {}^{(5)}R_L-2 {}^{(5)}\Lambda_L) +
 {}^{(5)}L_{L \rm m} \right],
 \label{bkaction}
 \en
 $S_{R \rm bulk}$ can be written correspondingly,
 and
 \be S_{\rm brane}=\int d^4 x\sqrt{-{\rm det}(g)} \left[
{1\over\ {8 \pi G_5}} K^\pm + L_{\rm brane}(g_{\alpha\beta},\psi)
\right].
 \label{bnaction}
 \en
Here $G_5$ is the 5d gravitational constant, ${}^{(5)}R$ and
${}^{(5)}L_{\rm m}$ are the 5d scalar curvature and the matter
Lagrangian in the bulk, respectively. A quantity with subscript
$L$ denotes it is valued in the left bulk and $R$, the right bulk.
${}^{(5)}\Lambda$ is the cosmological constant in the bulk. $x^\mu
~(\mu=0,1,2,3)$ are the induced 4d coordinates on the brane,
$K^\pm$ is the trace of extrinsic curvature on either side of the
brane and $L_{\rm brane}(g_{\alpha\beta},\psi)$ is the effective
4d Lagrangian, which is given by a generic functional of the brane
metric $g_{\alpha\beta}$ and matter fields $\psi$.

%%%%%%%%%%
Consider the brane Lagrangian
 \be
  L_{\rm brane}=
 \frac{1}{16\pi G} (R-2\lambda) + L_{\rm m},
 \label{branel}
 \en
 where $\lambda$ is the cosmological constant on the brane, $L_{\rm m}$
 denotes matter confined to the brane and  $R, G$, the 4d scalar curvature and gravitational constant respectively.
 We assume that the 5d bulk space includes only a
cosmological constant ${}^{(5)}\Lambda$. It is just a generalized
version of the DGP model, which is obtained by
setting  $\lambda=0$ as well as ${}^{(5)}\Lambda=0$. The covariant
equations in the case of ${}^{(5)}\Lambda_L={}^{(5)}\Lambda_R$
have been obtained in \cite{eff}, for
${}^{(5)}\Lambda_L\neq{}^{(5)}\Lambda_R$, the covariant equations
were derived in \cite{noninduce}.

    A mirror symmetric closed brane-world instanton $M$ can
be constructed by excising the spacetime region at $r> r_0$ left
$M_R$ and gluing two copies of the remaining spacetime along the
4-hypersurface at $r=r_0$ , $M=M_R^z\bigcup M_R^y$. We consider a
more general class of instantons without mirror symmetry.
Certainly the junction condition (\ref{jun}) must be satisfied.
Furthermore the energy momentum tensor on the brane is constrained
by the second fundamental form of the brane relative to the two
sides of the bulk. The relation is Israel's junction condition

 \be
 [K_{\mu\nu}-Kg_{\mu\nu}]^{\pm} =8\pi G_5\tau_{\mu\nu},
 \label{israel}
 \en
 where $\tau_{\mu\nu}$ is the effective energy momentum stress tensor on the
 brane,  $K_{\mu\nu}$ is the second fundamental form of the brane,
  $K=g_{\mu\nu}K^{\mu\nu}$, $[K_{\mu\nu}-Kg_{\mu\nu}]^{\pm}=[K_{\mu\nu}-Kg_{\mu\nu}]^+
  - [K_{\mu\nu}-Kg_{\mu\nu}]^-$ and  $[K_{\mu\nu}-Kg_{\mu\nu}]^+$ or
  $[K_{\mu\nu}-Kg_{\mu\nu}]^-$ is the value of the expression at one
 side of the brane respectively.
 It is easy to get from (\ref{branel}),
 \be
 \tau_{\mu\nu}=-\frac{1}{8\pi G}(\lambda g_{\mu\nu}+G_{\mu\nu})
 -2 {\delta L_m \over \delta
g^{\mu\nu}}  +g_{\mu\nu}L_m.
 \label{em}
 \en

We omit matter term $L_m$ in this section for the sake of
instanton solution. From the lemma we have
 \be
 R_{\mu\nu}=3\epsilon b^{-2} g_{\mu\nu}.
 \label{definec}
 \en
  One knows that $b$ only depends on the fifth dimension in above
  equation. So in (\ref{em}) the induced term $G_{\mu\nu}$ just
  acts as an cosmological constant from the brane view.

    From (\ref{israel}) we arrive at
 \be
 \widetilde{K}{_{\mu\nu}}=g_{\mu\nu} x\left(-\epsilon b^{-2}+\frac{\lambda}{3}\right),
 \label{k}
  \en
 where $x=\frac{G_5}{G}$.

 Without announcement for a  quantity $Q$, $\overline{Q}=\frac{1}{2}
 (Q^++Q^-)$ , $ \widetilde{Q}=Q^+-Q^-$, $Q^+$ or $Q^-$ is the value of the quantity at one
 side of the brane respectively.
  On the other hand
 \be
 K_{\mu\nu}=\frac{1}{2}{\cal L}_{\vec{n}} g_{\mu\nu} =\frac{b'}{b}
 g_{\mu\nu}.
 \label{kother}
 \en
 If the brane stands at $r=r_1$ relative to the left bulk, then (\ref{k})
  and (\ref{kother}) give
 \be
 \widetilde{(\frac{b'}{b})}=x\left(-\epsilon
 b^{-2}+\frac{\lambda}{3}\right),
 \label{junction}
 \en
  which yields, for positive curvature brane,
  \be
   \lambda=3\left[
  \left(l_1\sinh (\frac{r_1}{l_1})\right)^{-2}+\frac{1}{x}\left(\frac{1}{l_1}\coth(r_1/l_1)+
  \frac{1}{l_2}\coth(r_2/l_2)\right)\right],
 \label{plambda}
  \en
 for negative curvature brane,
  \be
  \lambda=3\left[-
  \left(l_1\cosh (\frac{r_1}{l_1})\right)^{-2}+\frac{1}{x}\left(\frac{1}{l_1}\tanh(r_1/l_1)+
  \frac{1}{l_2}\tanh(r_2/l_2)\right)\right],
 \label{nlambda}
  \en
  and for Ricci flat brane
  \be
 \lambda=3\frac{1}{x}\left[\frac{1}{l_1}+
  \frac{1}{l_2}\right].
 \label{flambda}
  \en

\begin{figure}
\centering
\includegraphics[totalheight=3in]{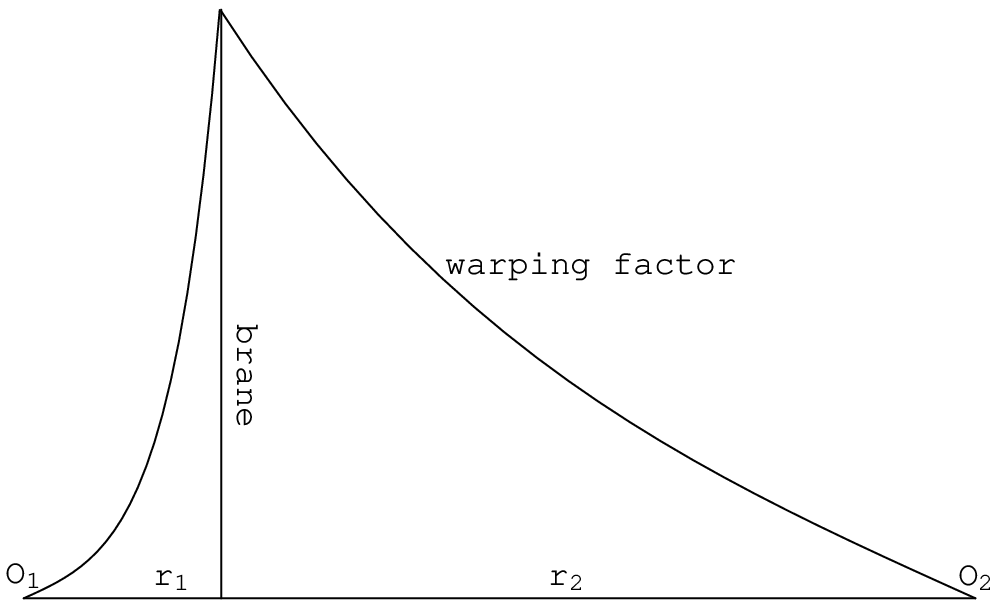}
\caption{ Warping factor across the two bulk.
\\In this figure $l_2=4l_1$. One can see from the figure that the warping factor
is not smooth at the position where brane stands, which provides
the energy momentum stress tensor on the brane. }
 \label{warp}
 \end{figure}

 Under these conditions: 1. the brane Lagrangian (\ref{branel}) does not contain induced
 gravity term $R$; 2. the instanton is mirror symmetric; and 3.
 the brane in the instanton is a positive curvature brane the junction
  condition (\ref{junction}) degenerates to the condition
 in \cite{garriga}. In
 fig.  \ref{warp} we plot the warping factor $b(r)$ across the bulk.
 For Ricci flat and negative cases the qualitative properties of
 the warping factors is the same as positive case.

\section{instanton action}
 In spirits of sum-over-history formulism path integral extends
 over all paths-differentiable and non-differentiable. we know
 that the measure is concentrated on non-differentiable paths
 in the ordinary path integral case. In flat space case we can smooth the
 non-differentiable path by general analytical techniques. On the other hand because of
 the singularity theorem any reasonable spacetime must consist of  one singularity at
 least. The discovery of cosmological constant, via breaking energy conditions, helps to escape the singularity theorem. But even
 in the case singular spacetime we can define the gravitational
 action in a meaningful way. In the case of black holes that the
 path integral should be taken over Euclidean, that is, positive definite metrics.
  This means that the
singularities of black holes, like the Schwarzschild solution, do
not appear on the Euclidean metrics which do not go inside the
horizon. Instead the horizon is like the origin of polar
coordinates. The action of the Euclidean metric is therefore well
defined. Based on these considerations we permit singular paths,
such as bulk which is Einstein space contain black strings. In the
following we calculate the actions of various paths. As the first
step we omit the induced gravity term in brane action
(\ref{bnaction}) for a while . So the Euclidean action of the
instanton

 \be
   S_E = {1\over 16\pi G_5}\left(\int d^4x \sqrt{{\rm det} g_E}
   (4\overline{K})+\sum_{L,R}\int d^5x \sqrt{{\rm det}({}^{(5)}g_E)}(2
  {}^{(5)} \Lambda-{{}^{(5)} R})\right)
 \nonumber
  \en
 \be
  +\frac{1}{16\pi G}\int
 d^4x\sqrt{{\rm det} g_E}
 (2\lambda-R)~~~~~~~~~~~~~~~~~~~~~~~~~~~
 \label{dsinads}
  \en
 becomes
 \be
   S_E = {1\over 16\pi G_5}\left(\int d^4x \sqrt{{\rm det} g_E}
   (4\overline{K})+\sum_{L,R}\int d^5x \sqrt{{\rm det}({}^{(5)}g_E)}(2
  {}^{(5)} \Lambda-{{}^{(5)} R})\right)
 \nonumber
  \en
 \be
  +\frac{1}{16\pi G}\int
 d^4x\sqrt{{\rm det} g_E}
 (2\lambda).~~~~~~~~~~~~~~~~~~~~~~~~~~~~~~~
 \label{dsinads1}
  \en
 In case of mirror symmetric instanton $K^+=-K^-,~\overline{K}=0$ then
 GH boundary term of the bulk vanishes. But in case of asymmetric instanton
 it is necessary to include such a term in the action.
 Using the junction condition (\ref{plambda})
 , the action of  positive curvature brane in the negative curvature Einstein bulk,

  \be
 S_E^+=\frac{V^+}{2 \pi G_5}W^+(l_1,l_2,r_1,r_2),
 \label{positiveac}
 \en
  where
  \be
  V^+=\int d^4x \sqrt{{\rm det} h_E}~,
  \en
 $h_E$ takes the positive curvature metric given in
 (\ref{mcbulk}), and thus $V^+$ is dimensionless and
 \be
 \begin{array}{lll}
 W^+(l_1,l_2,r_1,r_2) =  &-&\frac{l_1^4
 }{4l_2}\coth(r_2/l_2)\sinh^4(r_1/l_1)+
 \frac{3}{8}(l_1^2r_1+l_2^2r_2)
 \\ &-& \frac{1}{4}l_1^3 \sinh(2r_1/l_1)+\frac{1}{32}l_1^3\sinh(4r_1/l_1)
 \\ &-& \frac{3}{16}l_2^3\sinh(2r_2/l_2)
.
 \label{fourp}
 \end{array}
 \en
 By using (\ref{warpj}) one can eliminates a parameter, for example $r_2$, from
 (\ref{fourp}), then we obtain
 \be
 \begin{array}{llll}
 W^+(l_1,l_2,r_1)&=&\frac{1}{32} [12l_1^2r_1+12l_2^3
 arcsh(\frac{l_1}{l_2}\sinh(r_1/l_1)) \\ &+&16l_1^3\sinh^3(r_1/l_1)
 \sqrt{1+ (\frac{l_1}{l_2}\sinh(r_1/l_1))^2}-
 8l_1^3\sinh(2r_1/l_1)+l_1^3 \sinh(4r_1/l_1) \\ &-&
 8l_2^3\sinh(2arcsh(\frac{l_1}{l_2}\sinh(r_1/l_1)))
 +l_2^3\sinh(4arcsh(\frac{l_1}{l_2}\sinh(r_1/l_1))) ].
 \label{plus3p}
 \end{array}
 \en
 One can check the result in \cite{garriga} is a special case of
 our result (\ref{plus3p}) under the mirror symmetry condition
 $l_1=l_2$.

  Just by the same method we derive the action of instanton with negative curvature
 brane,

 \be
 S_E^-=\frac{V^-}{2 \pi G_5}W^-(l_1,l_2,r_1),
  \en
  where
  \be
  V^-=\int d^4x \sqrt{{\rm det} h_E}~,
  \en
 $h_E$ takes the negative curvature metric given in
 (\ref{mcbulk}), $V^-$ is dimensionless and
 \be
 \begin{array}{lll}
 W^+(l_1,l_2,r_1) =&-&\frac{1}{2}l_1^3\cosh^3( r_1/l_1)
 (1+\frac{l_1}{l_2}\cosh(r_1/l_1))\sqrt{1-\frac{2l_2}{l_2+l_1\cosh(r_1/l_1)}} \\
 &+&\frac{1}{32} [l_1^2(12r_1+8l_1\sinh(2r_1/l_1)
 +l_1\sinh(4r_1/l_1))+l_2^3(12arcsh(\frac{l_1}{l_2}\cosh(r_1/l_1))\\
 &+& 8\sinh(2arcch(\frac{l_1}{l_2}\cosh(r_1/l_1)))+
 \sinh(4arcch(\frac{l_1}{l_2}\cosh(r_1/l_1))))] .

 \label{neg3p}
 \end{array}
 \en

  The action of the instanton with Ricci flat brane can be obtained by
  analogy ,
   \be
 S_E^{flat}=\frac{V^{flat}}{2 \pi G_5}W^{flat}(l_1,l_2,r_1),
 \label{aflat}
 \en
  where
  \be
  V^{flat}=\int d^4x \sqrt{{\rm det} h_E}~,
  \en
 $h_E$ takes the Ricci flat metric given in
 (\ref{mcbulk}), $V^{flat}$ is also dimensionless and
 \be
  W^{flat}(l_1,l_2,r_1)=-\frac{1}{4}\left[l_1^3(1-e^{4r_1/l_1})+(1+e^{4r_1/l_1})l_2^3\right].
 \label{aflat}
  \en

 We know some Euclidean spaces whose 4-volumes are well defined.
 First for positive curvature
Einstein branes there are two examples whose volumes we can
calculate---de Sitter space and Nariai space. For a de Sitter
brane
 \be
 V^+(dS)=\frac{8}{3}\pi ^2.
 \label{v4ds}
 \en

\begin{figure}
\centering
\includegraphics[totalheight=3in]{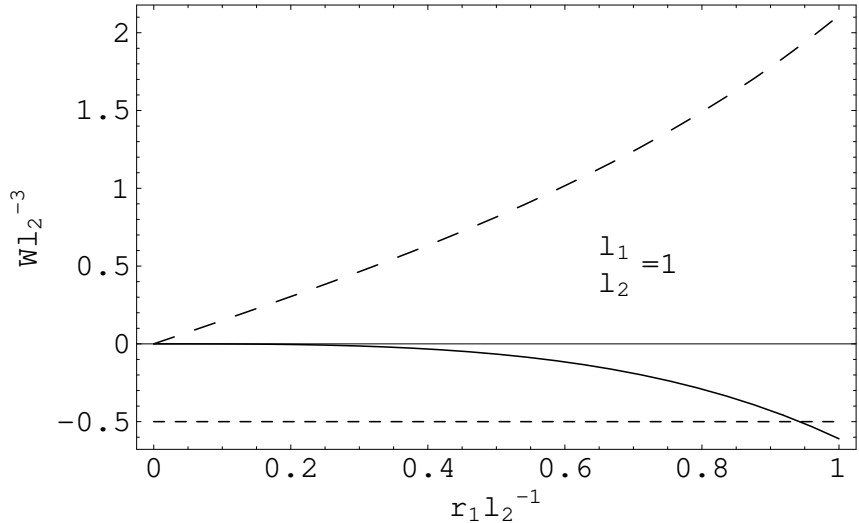}
\caption{ Actions of per unit 4-volume of instantons with mirror
symmetry. The solid curve denotes $W^- /l_2^3$, $W^- /l_2^3$ dwells
on the long dashing curve and $W^{flat} /l_2^3$ resides on the short
dashing curve. }
 \label{pfm11}
 \end{figure}

\begin{figure}
\centering
\includegraphics[totalheight=3in]{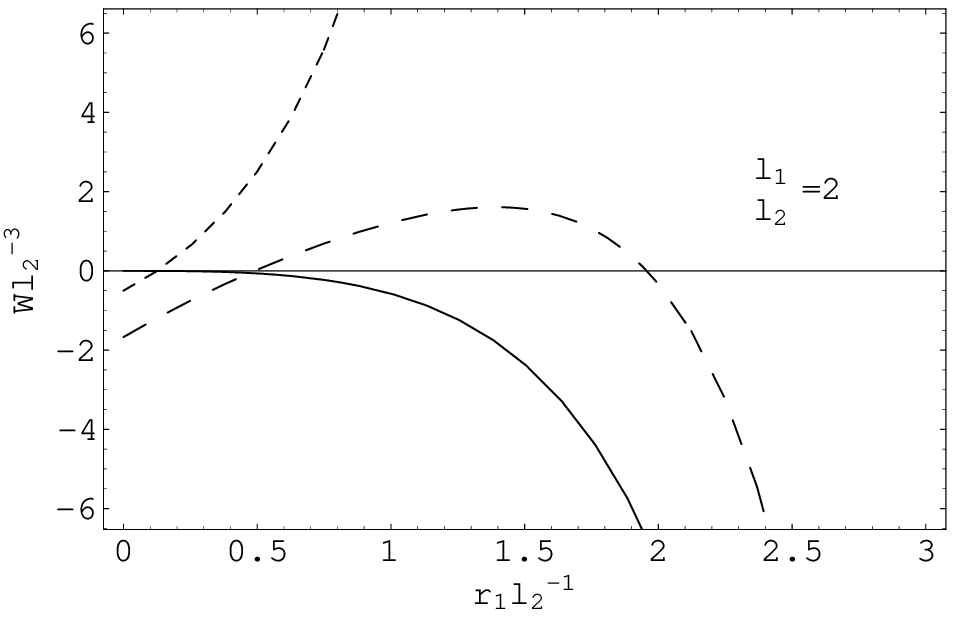}
\caption{ Actions  per unit 4-volume of instantons without mirror
symmetry, in which $\frac{l_1}{l_2}=2$, $l_2=1$. The solid curve
denotes $W^- /l_2^3$, $W^- /l_2^3$ dwells on the long dashing curve
and $W^{flat} /l_2^3$ resides on the short dashing curve.}
 \label{pfm21}
 \end{figure}

\begin{figure}
\centering
\includegraphics[totalheight=3in]{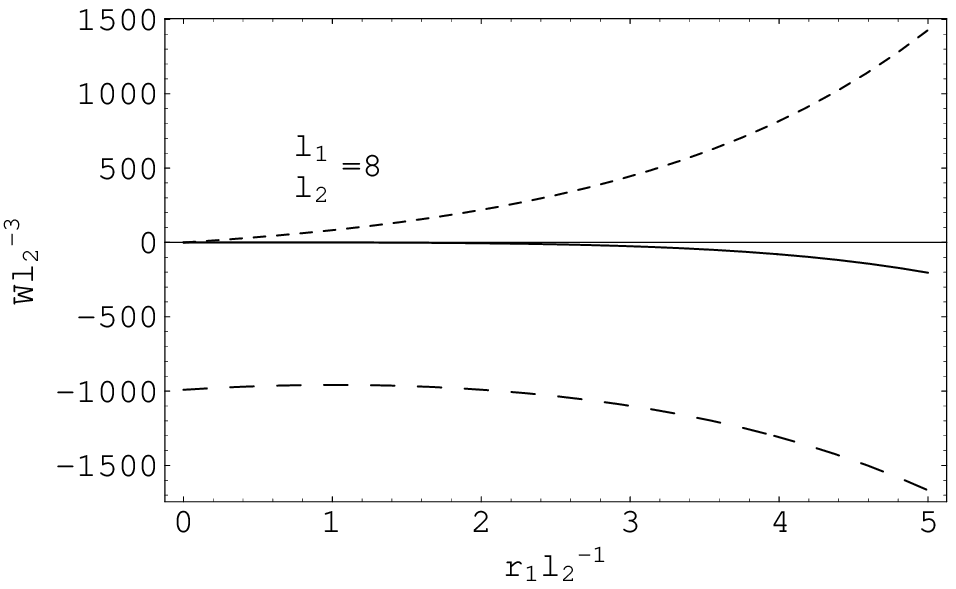}
\caption{ Actions  per unit 4-volume of instantons without mirror
symmetry, in which $\frac{l_1}{l_2}=8$, $l_2=1$. The solid curve
denotes $W^- /l_2^3$, $W^- /l_2^3$ dwells on the long dashing curve
and $W^{flat} /l_2^3$ resides on the short dashing curve.}
 \label{pfm81}
 \end{figure}

 For a Nariai brane we have $m=\frac{1}{3\sqrt{3}}$ in
 (\ref{sch(a)ds}) and the topology of the brane becomes $S^2\times
 S^2$. Hereby
 \be
 V^+(Na)=\frac{16}{9}\pi^2.
 \label{v4na}
 \en
 Then for Ricci flat Einstein branes we also presents some examples.
 One  knows  the volume of RS Euclidean (flat)
 brane is divergent without
 proper identity. But as we discussed above one can introduce
 instantons containing singularities. We calculate the actions
 of 4d Schwarzschild and Kerr metric here. Under these conditions
 we have to add a GH boundary term of the brane (boundary of
 boundary) in action (\ref{dsinads1}). Considering this term
 , (\ref{aflat}) becomes
 \be
 S_E^{flat}=\frac{V^{flat}}{2 \pi G_5}W^{flat}(l_1,l_2,r_1)+
 \frac{1}{8\pi G}\int d^3x I,
  \label{boundary}
 \en
 where $I$ is the trace of the second fundamental form of the
 boundary of the brane. For Schwarzschild solution the extra part
 of the
 action is $4\pi G m^2$, therefore
 \be
 S_E^{flat}=\frac{V^{flat}}{2 \pi G_5}W^{flat}(l_1,l_2,r_1)+
 4\pi G m^2.
 \en
 Here $m$ is the mass of the black hole. Note that the dimension
 $m$ is [mass] in the above equation which is different from $m$
 in (\ref{sch(a)ds}), where $m$ is dimensionless. One knows that
 for Kerr solution the extra part of the action is \cite{GH}     $$
 2\pi m \frac{r_+^2+J^2m^{-2}}{r_+-r_-}, $$
 where $m$ is the black hole mass, $r_+$ is the radius of outer
 horizon, $r_-$ is the inner horizon and $J$ is the angular
 momentum of the black hole. There is another point to explain.
 One knows the schwarzschild black hole on the brane is an
 extensive object which is a black string in the 5d  bulk. We only
 obtain the boundary term on brane, but how does the boundary of the
 string act off the brane? Generally speaking temperature makes
 no sense on an Einstein manifold with negative curvature.  In fact the total actions of the bulk have been
 included in the first term in (\ref{boundary}). However we still do
 not find the 4-volume of the brane appearing in (\ref{boundary}) and
 then the concrete value of the action $S_E^{flat}$ is left open.
 Nor do we know the 4-volume of the negative curvature brane
 without any compactifications ``by hand". Whereas in a sense
 one can compare the actions of per unit 4-volume
 all the same. We draw actions per unit 4-volume
 $W^+,~W^{flat},~W^-$ in figs. \ref{pfm21} and \ref{pfm81}, where
 the two instantons are in different asymmetric degrees.  In order to contrast
 with symmetric case we also draw $W^+,~W^{flat},~W^-$ of mirror
 symmetric instantons in fig. \ref{pfm11}.

\begin{figure}
\centering
\includegraphics[totalheight=3in]{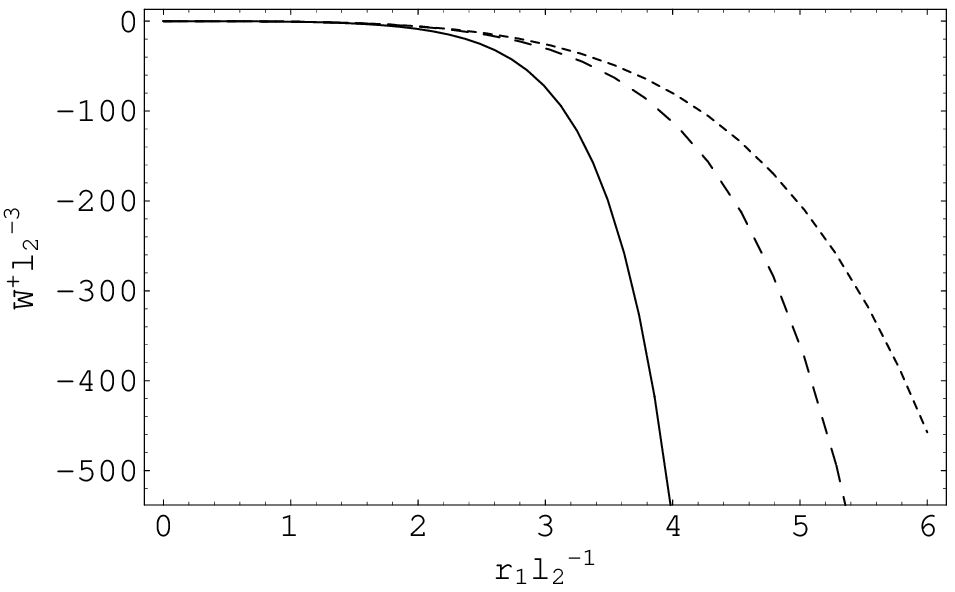}
\caption{ Actions per 4-volume of positive brane instantons with a
fixed half of bulk and a variable half of bulk. In this figure the
right bulk with $l_2=1$ is fixed and the left bulk with
$l_1/l_2=1,~ l_1/l_2=4,~l_1/l_2=8$ respectively. The solid line
denotes the left bulk of $l_1/l_2=1$, $l_1/l_2=4$ dwells on the
long dashing line and the left bulk of $l_1/l_2=8$ resides on the
short dashing line. As we expected the absolute value of the
action decreases when $r_1$ increases.}
 \label{ppp}
 \end{figure}

\begin{figure}
\centering
\includegraphics[totalheight=3in]{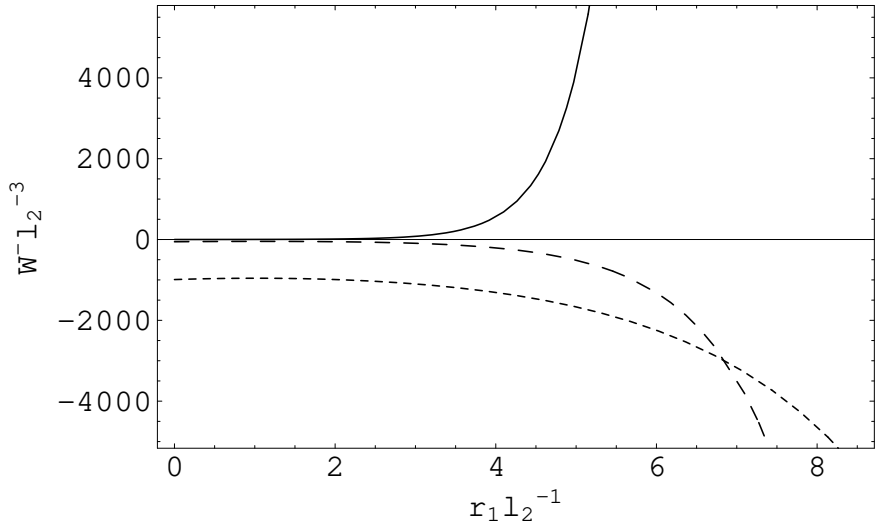}
\caption{ Actions per 4-volume of positive brane instantons with a
fixed half of bulk and a variable half of bulk. In this figure the
right bulk with $l_2=1$ is fixed and the left bulk with
$l_1/l_2=1,~ l_1/l_2=4,~l_1/l_2=8$ respectively. One finds
properties interestingly---the symmetric instanton always gets the
biggest action (the least creation probability, if probability
here is well defined). When $r_1$ is small the instanton
$l_1/l_2=4$ get the least action but with increasing of $r_1$ the
most the action asymmetric instanton $l_1/l_2=8$ soon becomes the
smallest among the three.}
 \label{mmm}
 \end{figure}

\begin{figure}
\centering
\includegraphics[totalheight=3in]{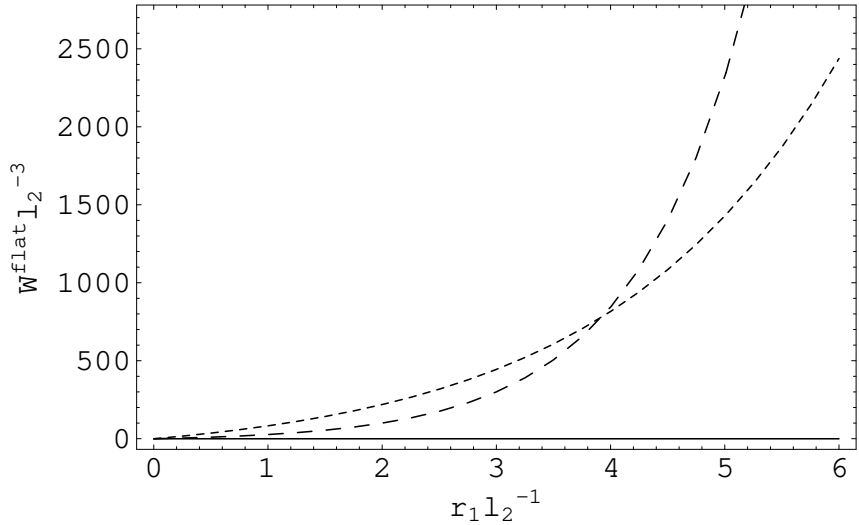}
\caption{Actions per 4-volume of positive brane instantons with a
fixed half of bulk and a variable half of bulk. In this figure the
right bulk with $l_2=1$ is fixed and the left bulk with
$l_1/l_2=1,~ l_1/l_2=4,~l_1/l_2=8$ respectively. The convention
follows the above figure. From (\ref{aflat}) we know $W^f$ is
exactly a constant in case of mirror symmetric instanton, as shown
in the figure, and it always gets the least value.}
 \label{fff}
 \end{figure}

Generally for mirror symmetric instanton the whole instanton is
fixed if we fix the bulk on one side of the brane, that is, the
bulk on the other side can be obtained by reflection and the
energy momentum tensor on the brane is just the difference between
the second forms along two sides of the brane. However for the
asymmetric case fixing one half of the bulk is not enough to fix
the whole instanton. We have many choices of other halves of bulk
and the corresponding branes. It is interesting to study the
actions of such a sequence of instantons. We present our
 results in fig. \ref{ppp}, \ref{mmm} and \ref{fff}.

   Now we turn to the action with an induced Ricci term
   (\ref{dsinads}). Integrate it straightforwardly
   \be
   S_E=-\left[\frac{1}{12\pi G_5}\sum_{L,R}\int drb^4(r){}^{(5)}\Lambda+
   \frac{\lambda b^4(r_1)}{24\pi G}+ \frac{b^6(r_1) \epsilon}{4\pi G}
   \right]V~,
   \label{induce}
  \en
 where $\epsilon$ is defined by (\ref{normalc}), $V$ is the volume of the
 4-Euclidean-Einstein brane with the metric $h_E=ds_4^2$ in (\ref{mcbulk}) and
 $V$ is a dimensionless number.
 The final integration in (\ref{induce})
 is fairly easy  while the final result is rather messy. But paying our attention to (\ref{em}) and
 (\ref{definec})
 we see that on an Einstein brane the effect of the induced Ricci tensor
 acts as a cosmological constant on the brane. We find the
 final result of (\ref{induce}) behaves similar to the case of
 the action without induced term (\ref{dsinads1})
 so it sheds no more light on our understanding of this
 asymmetric brean creation problem.

\section{conclusions and discussions}

 We present a quantum cosmological scenario of brane world creation
  with an induced gravity term on the brane.
 In this scenario a brane is created together with bulk from nothing.
 The quantum creation is described by the brane instanton---a positive, negative
 or flat Einstein 4-manifold, which separates
two asymmetric patches of negative curvature Einstein 5-manifolds.
We study all three classes of branes residing in a negative
curvature Einstein manifold and find they dwell in different
positions, up to a boost isometry one another for different
classes. All the branes in the same classes can be written in
standard form simply by a rescaling of $radial$ coordinate. Then
we analyze the junction condition of brane with asymmetric bulk in
induced gravity frame.

  In Euclidean quantum gravity formulism, which is the application of
  quantum path integral formulism in gravity theory, we have
 \be
 p\propto e^{-2S_E},
 \label{su}
 \en
 where $p$ is the probability associated to the path. So the
 action of an instanton may offer clues of the creation
 probabilities of the instantons. As an example from (\ref{v4ds}), (\ref{v4na})
 , (\ref{positiveac}), and (\ref{su})
   one can immediately say the de Sitter brane is
 more possibly created than Nariai brane.
 We  investigate in detail the Euclidean action of three canonical
 types of instantons. We find that GH boundary term should be considered
 in the asymmetric case. We provide the analytical forms of the
 instanton by three parameters---the characteristic lengthes of the
 bulk on the left
 ,right and the position of the brane in instanton. We
 find for most brane metrics we are not so fortunate to obtain the
 definite action of an instanton. The total actions of open and flat brane are ill defined
 without proper identifyings put ``by hand", as shown in section IV.
 Even in the case of positive brane, for instance simple
 as Schwarzschild-de Sitter metric, we do not know its Euclidean action very well
 because it is a non equilibrium system, i. e., its temperature is not definite.
 So we compared the actions of the three
 types of instantons per unit 4-volume. We also present the
 actions of instantons with a brane gluing a fixed bulk but different
 other bulk. All the three canonical types of branes  are studied.

%\newpage

{\bf Acknowledgments:} We thank  R. G. Cai for warm helps
in preparation of this paper. Hongsheng Zhang  thanks J. Garriga and
M. Sasaki for discussions on their paper hep-th/9912118. Hongsheng
Zhang thanks Dawei Pang  for useful discussions. This work is supported by the Program for Professor of Special Appointment (Eastern Scholar) at Shanghai Institutions of Higher Learning, National Education Foundation of China under grant No. 200931271104, Shanghai Municipal Pujiang grant No. 10PJ1408100, and National Natural Science Foundation of China under Grant No. 11075106.


\begin{thebibliography}{99}



%%cite{10,Gribov:1977wm}
\bibitem{qc}
 G. Gibbons , Class. Quantum
 Grav., 15 ,2605(1998) .

 G. Gibbons  and J. Hartle ,  Physical Rev. D,
42 (1990), 2458-2468.
 %Wave function of the universe,

 J. Hartle  and S. Hawking,
 Physical Rev. D, 28, 2960(1983) .

 A. Vilenkin, Phys. Rev. D 30, 509 (1984),
Phys. Rev. D 33, 3560 (1986).
\bibitem{braneworld}
L. Randall and R. Sundrum, hep-th/9905221.

L. Randall and R. Sundrum,
%``An alternative to compatification,''
hep-th/9906064.

 N. Arkani-Hamed, S. Dimopoulos and G. Dvali,
 Phys. Lett. B429
,263(1998),
 hep-ph/ 9803315.

 I. Antoniadis, N. Arkani-Hamed, S. Dimopoulos and G. Dvali, Phys.
Lett. B436 (1998) 257, hep-ph/9804398.

\bibitem{garriga}
  J Garriga and M Sasaki, Phys.Rev. D62 ,043523(2000) ,
  hep-th/9912118.


\bibitem{akm}
 K. Aoyanagi and K. Maeda,
 %Creation of a brane world with Gauss-Bonnet term,
 Phys.Rev. D70, 123506(2004) , hep-th/0408008.

 S. Nojiri and S. Odintsov,hep-th 0409244.
\bibitem{fbrane}
  M.~Bianchi, A.~Collinucci and L.~Martucci,
  %``Magnetized E3-brane instantons in F-theory,''
  arXiv:1107.3732 [hep-th]; M.~Cvetic, I.~Garcia-Etxebarria and R.~Richter,
  %``Branes and instantons at angles and the F-theory lift of O(1) instantons,''
  AIP Conf.\ Proc.\  {\bf 1200}, 246 (2010)
  [arXiv:0911.0012 [hep-th]].
\bibitem{witten}
  E.~Witten,
  %``Branes, Instantons, And Taub-NUT Spaces,''
  JHEP {\bf 0906} (2009) 067
  [arXiv:0902.0948 [hep-th]].
\bibitem{angle}
  M.~Cvetic, I.~Garcia-Etxebarria and R.~2.~Richter,
  %``Branes and instantons intersecting at angles,''
  JHEP {\bf 1001} (2010) 005
  [arXiv:0905.1694 [hep-th]].



 \bibitem{brcrea}
 M.~-x.~Luo, S.~Zheng,
  %``E2 Instanton Effects and Higgs Physics In Intersecting Brane Models,''

  [arXiv:0804.3265 [hep-th]].



 D.~Forcella, I.~Garcia-Etxebarria, A.~Uranga,
  %``E3-brane instantons and baryonic operators for D3-branes on toric singularities,''
  JHEP {\bf 0903}, 041 (2009).
  [arXiv:0806.2291 [hep-th]].


 S.~Matsuura,
  %``Instanton Counting and Dielectric Branes,''
  JHEP {\bf 0809}, 083 (2008).
  [arXiv:0808.3493 [hep-th]].


 S. Nojiri and S. Odintsov,  JHEP 0112, 033(2001) , hep-th/0107134.

 Y. Chen and J. Lu, hep-th/0405265.


 M. Bouhmadi-Lopez, P. Gonzalezdiaz, A. Zhuk, Class.Quant.Grav. 19, 4863(2002)
 , hep-th/0208226.

  E. Lima, H. Lu, B. Ovrut, C. Pope, Nucl.Phys. B569, 247(2000)
,hep-th/9903001.

 P. Ho and Y. Wu, Phys.Lett. B420, 43(1998), hep-th/9708137.

 B. de Carlos, J. Roberts, Y. Schmohe, hep-th/0406171.




 \bibitem{non1}
P. Kraus, J. High Energy Phys. 9912, 011 (1999).

D. Ida, J. High Energy Phys. 0009, 014 (2000).

\bibitem{non2}

N. Deruelle and T. Dole¡¦zel, Phys. Rev. D 62, 103502 (2000).

 W.B. Perkins, Physics Lett. B 504, 28 (2001).

\bibitem{nonboth}
P Bowcock, C Charmousis, and R Gregory, Class. Quantum Grav. 17,
4745 (2000).

B. Carter and J.-P. Uzan, Nucl. Phys. B 606, 45 (2001).

 H. Stoica, H. Tye, and I. Wasserman, Phys. Lett. B 482, 205
(2000).

    L. A. Gergely, Phys.Rev. D68, 124011(2003) , gr-qc/0308072.



\bibitem{referee}
J.D. Brown, C. Teitelboim, Nucl.Phys.B297:787-836(1988).

    \bibitem{imbed}

 %Embedding Theorems and Higher Dimensional Physics,
    S. Seahra, gr-qc/0302015.
  %Global Embedding of Analytic Branes in Ricci-Flat MD Bulk Cosmology: Existence and Homotopy Results

    N. Katzourakis, math.DG/0411630.

\bibitem{Dvali}
G. Dvali, G. Gabadadze, M. Porrati,
 %``4D Gravity on a Brane in 5D Minkowski Space''
 hep-th/0005016.

 G. Dvali and G. Gabadadze, Phys. Rev.
D63, 065007 (2001).
\bibitem{eff}
K. Maeda, S. Mizuno and T. Torii,
%``effective brane equation,''
 gr-qc/0303039;  R.~-G.~Cai, H.~-S.~Zhang, A.~Wang,
  %``Crossing w=-1 in Gauss-Bonnet brane world with induced gravity,''
  Commun.\ Theor.\ Phys.\  {\bf 44}, 948 (2005).
  [hep-th/0505186].


 \bibitem{noninduce}
 L. Gergely, R. Maartens, gr-qc/0411097.
 \bibitem{GH}
G. Gibbons and S. Hawking, Phys. Rev. D15, 2752 (1977).

\end{thebibliography}
\end{document}